# Air pollution studies in terms of $PM_{2.5}$, $PM_{2.5-10}$, $PM_{10}$, lead and black carbon in urban areas of Antananarivo - Madagascar

E. O. Rasoazanany, N. N. Andriamahenina, H. N. Ravoson, Raoelina Andriambololona,
L. V. Randriamanivo, H. Ramaherison, H. Ahmed, M. Harinoely
*Institut National des Sciences et Techniques Nucléaires (Madagascar-INSTN)*
*B.P. 4279 - 101 Antananarivo, MADAGASCAR*

Atmospheric aerosols or particulate matters are chemically complex and dynamic mixtures of solid and liquid particles. Sources of particulate matters include both natural and anthropogenic processes. The present work consists in determining the concentrations of existing elements in the aerosols collected in Andravoahangy and in Ambodin'Isotry in Antananarivo city (Madagascar). The size distribution of these elements and their main sources are also studied. The Total Reflection X-Ray Fluorescence spectrometer is used for the qualitative and quantitative analyses. The results show that the concentrations of the airborne particulate matters $PM_{2.5-10}$ are higher than those of $PM_{2.5}$. The identified elements in the aerosol samples are Ti, Cr, Mn, Fe, Ni, Cu, Zn, Br, Sr and Pb. The average concentrations of these elements are also higher in the coarse particles than in the fine particles. The calculation of the enrichment factors by Mason's model shows that Cr, Ni, Cu, Zn, Br and Pb are of anthropogenic origins. The average concentrations of lead (2.8 ng.m$^{-3}$, 31.3 ng.m$^{-3}$ and 19.6 ng.m$^{-3}$ respectively in aerosols collected in Andravoahangy in 2007 and in 2008 and in Ambodin'Isotry in 2008) are largely lower than the average concentration of 1.8 µg.m$^{-3}$ obtained in 2000 in the Antananarivo urban areas. The concentration of black carbon is higher in the fine particles. The Air Quality Index category is variable in the two sites.

## 1. INTRODUCTION

Air pollution is a serious global issue, which threatens both human health and environment. The air pollutants criteria are airborne particulate matter (APM), ozone, carbon monoxide, sulfur dioxide, nitrogen dioxide and lead [1]. Over the last few years, there has been growing attention to black carbon and its potential impact on global warming. The concentrations of air pollutants may be affected by meteorological factors such as wind [2], atmospheric turbulence, stability and unstability of atmospheric temperature [3], and physical factors. Since 1996, the "Institut National des Sciences et Techniques Nucléaires (Madagascar-INSTN)" has been dealing with air pollution, especially pollution by lead and particulate matters in the city of Antananarivo. Studies showed that the mean concentration of lead (1.8 µg.m$^{-3}$) in Antananarivo urban areas is largely higher than the World Health Organization (WHO) adopted authorized value which is 0.5 µg.m$^{-3}$ [4]. The same problem occurred for the APM. As a follow-up of the previous studies, this work consists in determining the concentrations of black carbon. The black carbon is a component of particulate matter, or soot, produced from incomplete combustion of fossil fuel, biofuels and biomass.

## 2. MATERIALS AND METHODS

### 2.1. Aerosol sampling equipment and sample collection

Samplings of APM are done during 24 hours (Monday to Sunday) in two polluted locations (Andravoahangy and Ambodin'Isotry) in Antananarivo, the capital of Madagascar. These sites are selected for convenience and safety reasons. The sampling campaigns are divided into two parts: from February until March 2007 in Andravoahangy and from April until August 2008 in Andravoahangy and Ambodin'Isotry. The collections of APM are carried out using the samples collector which complies with the GENT $PM_{10}$ Stacked Filter Unit [5] provided by the International Atomic Energy Agency (IAEA, Vienna-Austria). The GENT air sampler is able to separate the particles having aerodynamic diameter lower than 2.5 µm known as fine particles ($PM_{2.5}$) and $PM_{2.5-10}$ with aerodynamic diameter ranging between 2.5 and 10 µm called coarse particles. The sampler is placed with its $PM_{10}$ inlet at about 7 m above the ground surface and operates with an air flow rate of 18 liters per minute. The samples are collected on two types of Nuclepore polycarbonate filters of diameter 47 mm respectively to collect the coarse fraction $PM_{2.5-10}$ (porosity 8 µm) and fine fraction $PM_{2.5}$ (porosity 0.4 µm) [5]. The filters are weighed with a microbalance before and after sampling to determine the mass of aerosols deposited on the filters.

## 2.2. Preparation of aerosol samples

The filters containing the aerosol samples are dissolved in solution in a digestion bomb in Teflon with mixtures of 500 µL of $CHCl_3$, 2 mL of $HNO_3$ and 1 mL of $H_2O_2$. Then the bomb is completely closed in order to reach a temperature of 165 °C in an oven for 4 hours [6]. In a small polyethylene bottle containing 995 µL of the latter solution, 5 g of yttrium (1 000 mg.L$^{-1}$) as internal standard are added.

## 2.3. Analysis Protocol

The analysis of aerosol samples is performed in Madagascar-INSTN laboratory. The total reflection X Ray Fluorescence (TXRF) method is used to measure the aerosol samples and to determine the elements in particulate matter collected on the filters. The TXRF is originally suggested as an analytical method by Yoneda and Horiuchi in 1971 to reduce the noise by placing the samples on an optically flat surface [7]. It is then developed by Aiginger and Wobrauschek in 1974. Thus, the analysis by total reflection X Ray Fluorescence is an improvement with regards to the dispersive energy X Ray Fluorescence (EDXRF).

## 3. RESULTS AND DISCUSSION

## 3.1. Airborne Particulate Matters ($PM_{2.5}$, $PM_{2.5-10}$ and $PM_{10}$)

The mean daily APM concentrations are higher than the 2005 WHO values set at 25 µg.m$^{-3}$ for $PM_{2.5}$ [8] in 21 out of 22 samples collected in Andravoahangy in 2007. For the particles which have an aerodynamic diameter smaller than 10 µm ($PM_{10}$), all the samples have APM concentrations exceeding the WHO and European Union (EU) guidelines values, set at 50 µg.m$^{-3}$ [9]. According to the United States Environmental Protection Agency (US EPA) standards [10], set at 35 µg.m$^{-3}$ for $PM_{2.5}$, 19 out of 22 samples collected in Andravoahangy in 2007 have APM concentrations exceeding the guidelines values. For the samples taken in Andravoahangy and Ambodin'Isotry in 2008, the same trend has been observed for $PM_{2.5}$ and $PM_{10}$. In conclusion, the air quality in the two sites does not comply with the standards values. Moreover, the particulate matters smaller than 10 µm in the two sampling sites are more concentrated in coarse particles than in the fine particles. This is likely due to the location of the sampling sites, which are close to the main roads, where atmospheric turbulence of dust particles produced by the daytime running cars is abundant. Figure 1 illustrates the concentrations of the APM in the aerosol samples taken in Andravoahangy in 2007.

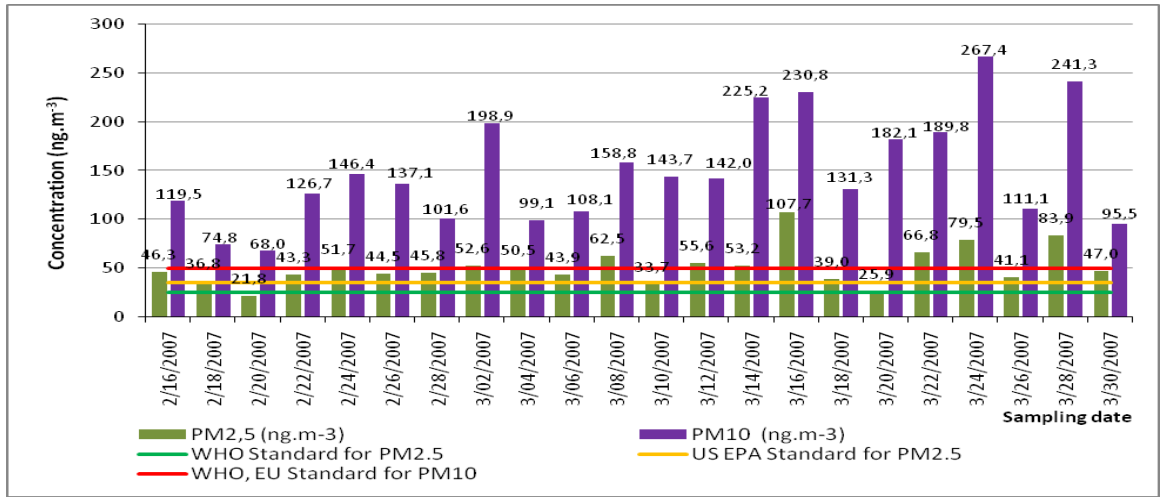

Figure 1. Daily average concentrations of $PM_{2.5}$ and $PM_{10}$ in the aerosol samples colleted in Andravoahangy in 2007.

### 3.2. Air Quality Index category and Health effects statements

The purpose of the Air Quality Index (AQI) is to help understand what local air quality means to our health. Concerning the AQI category adopted by US EPA regarding the concentrations of $PM_{2.5}$ and $PM_{10}$, the level of pollution can be referenced according to the value of the AQI. The most recent health effect information used with Air Quality Index (AQI) is pollutant-specific [11].

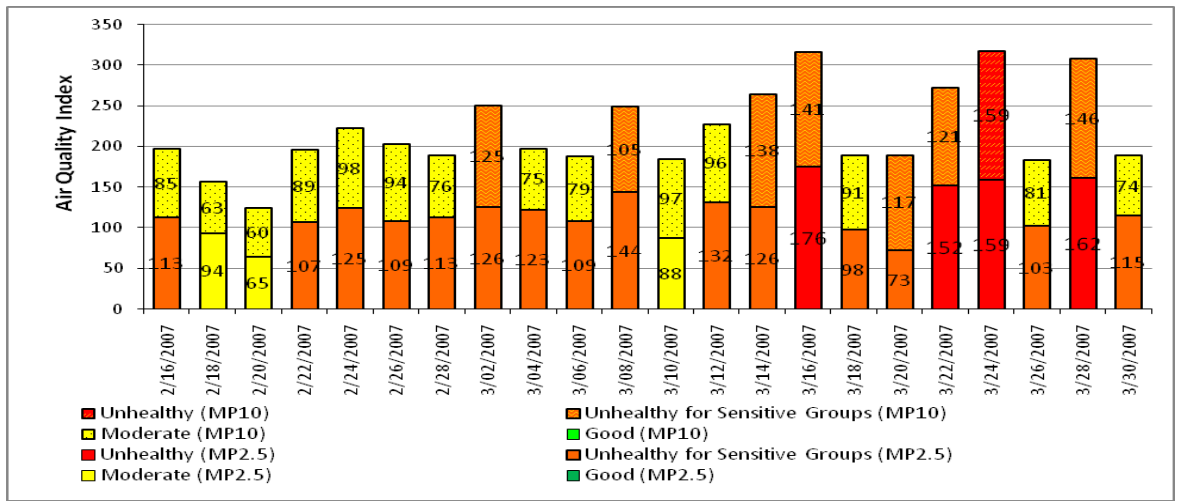

Figure 2. Daily average concentrations of $PM_{2.5}$ and $PM_{10}$ in the aerosol samples colleted in Andravoahangy in 2007.

In Andravoahangy, the air quality varies almost on daily basis (Figure 2). However, in 2007, most of the time (13 out of 22 days), the air quality falls within the "Unhealthy for Sensitive Groups" due to $PM_{2.5}$. The maximum value of AQI ranging between 101 to 150 is equal to 144. People with respiratory or heart disease, the elderly and children are most vulnerable. Therefore, children should limit prolonged exertion. The air quality on Friday 16[th] March, 2007; Thursday 22[nd] March, 2007; Saturday 24[th] March, 2007 and Wednesday 28[th] March, 2007 is qualified "Unhealthy" resulting from $PM_{2.5}$ with index values respectively equal to 176, 152, 159 and 162. As a consequence of $PM_{10}$ results in 2007, 14 out of 22 days, the air quality falls within "Moderate" (AQI between 51 to 100). The air quality can be thus qualified as acceptable. In 2008, the air quality is "Good" on Sundays, according to $PM_{10}$ results. The air quality can be qualified as satisfactory and poses little or no health risk. As for Ambodin'Isotry case in 2008, the air quality is almost "Moderate" according to $PM_{2.5}$ and $PM_{10}$ results. The maximum value of AQI is 98.

### 3.3. Elemental composition

The aerosol samples having aerodynamic diameter lower than 10 µm sampled the during day and night times (Monday to Sunday) contain ten elements like like titanium (Ti), chromium (Cr), manganese (Mn ), iron (Fe), nickel (Ni), copper (Cu), zinc (Zn), bromine (Br), strontium (Sr) and lead (Pb). Fe is the major component of the aerosol samples. Ti and Zn are minor elements whereas Cr, Mn, Ni, Cu, Br, Sr and Pb are trace elements. This study deals also with the average concentrations calculation of these elements present in the fine particles $PM_{2.5}$ and in the coarse particles $PM_{2.5-10}$ during the sampling period. The average concentrations of these elements are higher in the coarse particles than those in the fine particle. Concerning the aerosol samples taken in Andravoahangy in 2007, the total average concentrations of all elements are relatively low.

The total average concentrations of lead in aerosols collected in Andravoahangy in 2007 ($2.8 \pm 0.2$ ng.m$^{-3}$), taken in Andravoahangy in 2008 ($31.3 \pm 1.8$ ng.m$^{-3}$) and sampled in Ambodin'Isotry ($19.6 \pm 2.3$ ng.m$^{-3}$) are largely lower than the average concentration of 1.8 µg.m$^{-3}$ which is obtained in Antananarivo urban areas in 2000. This is due to the decision of the Government, on 18 December 2002, to stop all unleaded gasoline imports [12].

### 3.4. ENRICHMENT FACTORS

Enrichment factors (EF) are used to estimate the contribution of a specific source to the concentrations of elements in an aerosol sample [13]. It is important to use the abundances source in the crust to calculate the EF. For XRF and TXRF methods, it is recommended to use titanium (Ti) for the reference element [5]. According to Mason's model [14], the calculation of the enrichment factor is performed for all elements taking into account the concentration of these elements in the crust. Cr, Cu, Zn, Br and Pb are not derived solely from natural sources (soil, crust ...) but mostly from anthropogenic sources. The results show also that bromine is highly enriched in the aerosols collected in Andravoahangy and Ambodin'Isotry during the period of sampling (April to August 2008). The enrichment factors in two sites, calculated during 24 hours, are respectively equal to 177 and 471 for fine particles and 102 and 183 for coarse particles.

### 3.5. BLACK CARBON CONCENTRATIONS

Black (elemental) carbon (BC) is of special interest because it absorbs sun radiations, warms the air, and contributes to BC emissions, a product of incomplete combustion from coal, diesel engines [15]. There are several methods of estimating elemental carbon concentration. One of them is by using a smoke stain reflectometer. The smoke stain EEL Model 43D Reflectometer is an instrument that is able to provide an estimation of the black carbon concentration in collected air particulate matter on air filter samples by reflectance technique. This is carried out by measuring the darkness of the smoke stain obtained from the sampler. The calculation of the concentrations of the black carbon by Cohen's model shows that the greatest concern is for smaller "fine" particles ($PM_{2.5}$), which include almost all combustion related particles. The maximum value is 9.62. Elemental carbon is therefore a major component of fine urban aerosols [16]. Diesel vehicles are mostly responsible for black carbon emissions. However, we note also that the concentrations of black carbon are very low in the aerosols collected in Andravoahangy from February to March 2007.

## 4. CONCLUSION

The average concentrations of elements in the coarse particles $PM_{2.5-10}$ are greater than those in fine particles. This is due to daytime human activities. The total average lead concentrations in aerosols collected in Andravoahangy in 2007 ($2.8 \pm 0.2$ ng.m$^{-3}$), and in 2008 ($31.3 \pm 1.8$ ng.m$^{-3}$) and in aerosols sampled in Ambodin'Isotry ($19.6 \pm 2.3$ ng.m$^{-3}$) are lower than the mean concentration ($1.8$ µg.m$^{-3}$) obtained in Antananarivo urban areas in 2000. The highest values of the concentrations of black carbon are found in the $PM_{2.5}$. The primary sources include emissions from diesel engines. In 2008, the air quality related to $PM_{10}$ on Sundays is "Good" in Andravoahangy.

### Ackowledgments

The authors would like to thank the International Atomic Energy Agency (IAEA, Vienna-Austria) for its technical support and all the staff of Madagascar-INSTN for their valuable cooperation through the RAF/4/019 project.

Warm thanks go to the two Directors of Public Elementary Schools of Andravoahangy and Ambodin'Isotry who have accepted to host the air samplers in their establishments during the sampling campaigns.